
\input harvmac

\nopagenumbers

\rightline{WISC-MILW-95-TH-10}
\vskip 0.7in
\centerline{\titlefont Semi-infinite Throat as the End-state Geometry}
\vskip 0.1in
\centerline{\titlefont of two-dimensional Black Hole Evaporation}

\vskip 0.5in
\qquad \qquad $\;$ {\bf Sukanta Bose},\footnote{$^{*}$}{
Electronic address: {\it bose@csd.uwm.edu}} \quad
{\bf Leonard Parker},\footnote{$^{**}$}{Electronic address:
{\it leonard@cosmos.phys.uwm.edu}} $\>$ and $\>$
{\bf Yoav Peleg}\footnote{$^{***}$}{Electronic address: {\it
yoav@alpha2.csd.uwm.edu}}
\vskip 0.08in \centerline{\it Department of Physics}
\centerline {\it   University of Wisconsin-Milwaukee, P.O.Box 413}
\centerline{\it   Milwaukee, Wisconsin 53201, USA }

\vskip 1in

\noindent
ABSTRACT: We study a modified two-dimensional dilaton gravity
theory which is exactly solvable in the semiclassical approximation
including back-reaction. The vacuum solutions of this modified theory
are asymptotically flat static space-times.
Infalling matter forms a
black hole if its energy is above a certain threshold. The black hole
singularity is initially hidden behind a timelike apparent horizon.
As the black hole evaporates by emitting Hawking radiation,
the singularity meets the shrinking
horizon in finite retarded time to become naked.
A natural boundary condition exists at the naked singularity such that for
general infalling matter-configurations
the evaporating black hole geometries can be matched continuously to
a unique static end-state geometry.
This end-state geometry is asymptotically
flat at its right spatial infinity, while its left spatial infinity is
a semi-infinite throat extending into the strong coupling region.

\Date{ }

\def\R{R^{(2)}}

%
\pageno=1

Hawking's discovery that black holes radiate thermally [{\ref\Hawking{S.W.
Hawking, {\it Comm. Math. Phys.} {\bf 43}, 199
(1975).},{\ref\ParkerWald{L. Parker,
{\it Phys. Rev.} {\bf D12}, 1519 (1975); R.M. Wald, {\it Comm. Math. Phys.}
{\bf 45}, 9 (1975).}},{\ref\Hawkingg{S.W. Hawking, {\it Phys. Rev.}
{\bf D14}, 2460 (1976).}}]
gave rise to a long-standing question concerning the consequences
of combining quantum theory and general
relativity. [{\ref\Hawkinggg{S.W. Hawking,
{\it Comm. Math. Phys.} {\bf 87}, 395 (1982).}},{\ref\thooft{G. 't Hooft,
{\it Nucl. Phys.} {\bf B256},
727 (1985) ; {\it Nucl. Phys.} {\bf B335} 138 (1990) ;
C.R. Stephens, G. 't Hooft and B.F. Whiting, {\it Class. Quan. Grav.}
{\bf 11}, 621 (1994).},{\ref\Susskind{L. Susskind, {\it Phys. Rev. Lett.}
{\bf 71}, 2367 (1993);
{\it Phys. Rev.} {\bf D49}, 6606 (1994); L. Susskind, L. Thorlacius and
J. Uglum, {\it Phys. Rev.} {\bf D48}, 3743 (1993).}},{\ref\Aharonov{Y.
Aharonov, A. Casher and S. Nussinov,
{\it Phys. Lett.} {\bf B191}, 51 (1987).}}]
Does evolution from an initial pure state take place unitarily to a
final pure state or non-unitarily to a final mixed state?
Intimately linked to this question is the final geometry resulting
from black hole evaporation.

Here we present a specific two-dimensional (2D) dilaton gravity model in
which a black hole evaporates leaving a static semi-infinite throat as the
end-state or ``remnant" geometry.
Our model is a modification of the CGHS
model.[{\ref\CGHS{C. Callan, S. Giddings,
J. Harvey and A. Strominger, {\it Phys. Rev.} {\bf D45}, R1005 (1992).}}]
We solve the semiclassical equations and get closed-form
expressions for the metric and dilaton field.

The classical 2D CGHS action {\CGHS} is
\eqn\classicalaction{
S_{c\ell} = {1\over 2\pi} \int d^{2}x \sqrt{-g}
\left[ e^{-2\phi} \left( \R + 4 (\nabla \phi)^{2} + 4 \lambda^2 \right)
- {1\over 2} \sum_{i=1}^{N} (\nabla f_{i})^{2} \right]
,}
where $\phi$ is the dilaton field, $\R$ is the 2D Ricci scalar,
$\lambda$ is a positive constant, $\nabla$ is the covariant derivative,
and the $f_{i}$
are $N$ matter (massless scalar) fields. The action
{\classicalaction} describes a 2D effective theory
in the throat region of a 4D almost extreme magnetically charged
black hole.[{\ref\Horowitz{D. Garfinkle, G. Horowitz and A.
Strominger, {\it Phys. Rev. Lett.} {\bf 67}, 3140
(1991).}},{\ref\yoav{Y. Peleg, {\it Mod. Phys. Lett.} {\bf A9}, 3137 (1994).}}]
It may also be regarded as a 2D arena in which some of the main
questions about black hole evaporation can be studied.
Among the classical solutions stemming from the action
{\classicalaction} are vacuum solutions,
static black hole solutions, and dynamical solutions describing
the formation of a black hole
by collapsing matter fields. For a review see Ref. [{\ref\Harvey{J. Harvey
and A. Strominger, ``Aspects of Quantum Black Holes",
in the Proceedings of the TASI Summer School 1992, Boulder, Colorado
(World scientific, 1993);
S.B. Giddings, ``Quantum Mechanics of Black Holes", hepth@xxx/9412138
(1994).}}].

To study one loop quantum corrections and back-reaction one can
use the trace-anomaly for massless scalar fields in
two dimensions, $\langle T^{\mu}_{\mu}\rangle =
{\hbar\over 24} \R$, and find the effective
action $S_{PL}$ for which $\langle T_{\mu \nu}\rangle =
-{2\pi \over \sqrt{-g}} {\delta \over \delta g^{\mu \nu}} S_{PL}$.
This is the Polyakov-Liouville
action [{\ref\Polyakov{A.M. Polyakov, {\it Phys. Lett.}
{\bf B103}, 207 (1981).}}]
\eqn\Polyakovaction{
S_{PL} = - {\hbar \over 96\pi}
\int d^{2}x \sqrt{-g(x)}
\int d^{2}x' \sqrt{-g(x')} \R(x) G(x,x') \R(x')
,}
where $G(x,x')$ is a Green function for
$\nabla^{2}$. Here we take the large $N$ limit, in which $\hbar$ goes to
zero while $N\hbar$ is held fixed. In that limit the quantum corrections
for the gravitational and dilaton fields are negligible,
and one need take into
account only the quantum corrections for the matter (scalar) fields.
The one-loop effective action is then $S_{(1)}=S_{c\ell} +
N S_{PL}$.
There are no known analytic solutions to this one-loop effective
theory,
though there are some numerical ones.[{\ref\Lowe{D. Lowe, {\it Phys. Rev.}
{\bf D47}, 2446 (1993); T. Piran and A. Strominger,
{\it Phys. Rev.} {\bf D48}, 4729 (1993).}}]
In order to find analytic solutions including semiclassical
corrections, one can modify the
action as in [{\ref\RST{J. Russo, L. Susskind and L. Thorlacius,
{\it Phys. Rev.}
{\bf D46}, 3444 (1992).}},{\ref\deAlwis{S.P. deAlwis, {\it Phys. Lett.}
{\bf B289}, 278 (1992); {\it Phys. Lett.} {\bf B300}, 330
(1993).}},{\ref\Bilal{A. Bilal
and C. Callan, {\it Nucl. Phys.} {\bf B394}, 73 (1993).}}].
Our approach is similar, in that we modify the
original CGHS action {\classicalaction} and find
analytic solutions to the modified equations including
back-reaction. However, our analytic solutions
yield {\it closed-form} expressions for the metric and dilaton field.
This allows us to fully analyze the solutions.

We add to the classical action {\classicalaction} a local covariant term
of one-loop order,
\eqn\corrections{
S_{_{\hbox{corr}}} = {N\hbar\over 24\pi} \int d^{2}x \sqrt{-g} \left(
(\nabla \phi)^{2} - \phi \R \right)
.}
Now the total modified action including the one-loop Polyakov-Liouville
term is
\eqn\modifiedaction{
S_{\hbox{mod}} = S_{c\ell} +
S_{_{\hbox{corr}}} + NS_{PL}
.}
Using null coordinates $z^{\pm}$ and conformal gauge $g_{++}=g_{--}=0$,
$g_{+-} = - {1\over 2} e^{2\rho}$ ($ds^{2}=-e^{2\rho}dz^{+}dz^{-}$),
the action {\modifiedaction} can be written in the form
\eqn\modifiedconformal{
S_{\hbox{mod}} = {1\over \pi} \int dz^+dz^- \left[ 2\partial_{-}(\phi-\rho)
\partial_{+} \left( e^{-2\phi} -{\kappa\over 2} (\phi-\rho) \right)
+ \lambda^{2} e^{2(\rho-\phi)} + {1\over 2} \sum_{i=1}^{N} \partial_{+} f_{i}
\partial_{-} f_{i} \right]
,}
where $\kappa = {N\hbar\over 12}$. [From the point of view of string
theory, the action {\modifiedconformal} with free fields
$X \equiv e^{-2 \phi}$ and $Y \equiv \phi - \rho$
(which are flat target space coordinates),
describes a conformal field theory with
tachyon and dilaton backgrounds $\hbox{T}=-4\lambda^{2}e^{-2Y}$ and
$\Phi = -2X + 2\kappa Y$.[{\ref\Giddings{
S. Giddings and A. Strominger, {\it Phys. Rev.} {\bf D47}, 2454 (1993).}}]]
The action {\modifiedconformal} is also invariant
under the transformation\footnote{$^{\dag}$}{Unlike
in the RST model, {\RST}, in this model
the transformation is {\it exactly} the same as in the classical case.}{\RST}
$\delta \phi = \delta \rho =
\epsilon e^{2\phi}$, with the conservation equation
$\partial_{\mu} \partial^{\mu} (\phi - \rho) = 0$.
We therefore can complete the gauge fixing by choosing the
``Kruskal coordinates," $x^{\pm}(z^{\pm})$,
in which $\phi(x^{+},x^{-}) = \rho(x^{+},x^{-})$. In this Kruskal gauge the
equations of motion derived from the modified action {\modifiedconformal}
are {\it exactly} the same as the classical ones
\eqn\phirhoequations{
\partial_{x^{+}} \partial_{x^{-}}
\left( e^{-2 \rho (x^{+},x^{-})} \right) =
\partial_{x^{+}} \partial_{x^{-}}
\left( e^{-2 \phi (x^{+},x^{-})} \right) = - \lambda^{2}
}
\eqn\fequation{
\partial_{x^{+}} \partial_{x^{-}} f_{i}(x^{+},x^{-}) = 0
,}
while the constraints get modified by
non-local terms $t_{\pm}(x^{\pm})$ arising from the Polyakov-Liouville action.
In conformal gauge, one can use the trace anomaly of $N$ massless scalar
fields $f_{i}$
to obtain $\langle T^{f}_{+-} \rangle = -\kappa \partial_{+}\partial_{-} \rho$
and integrate [{\ref\christensen{S.M. Christensen and
S.A. Fulling, {\it Phys. Rev.}, {\bf D15},
2088 (1977).}},{\ref\Parker{L. Parker, ``Aspects
of Quantum Field Theory in Curved Spacetime", in {\it Recent Developments
in Gravitation}, eds. S. Deser and M. Levi (Plenum, New York,
1979).}},{\ref\Birrell{N.D. Birrell and P.C.W. Davies, {\it Quantum
fields in curved space}, (Cambridge Univ. Press, Cambridge, 1982).}}]
the equation
$\nabla^{\mu} \langle T^{f}_{\mu \nu}\rangle = 0$ to get the quantum
corrections to the energy-momentum tensor of the $f_{i}$ matter fields
\eqn\nonlocalterms{
\langle T^{f}_{\pm\pm}\rangle = \kappa \left( \partial^{2}_{\pm} \rho
- (\partial_{\pm} \rho)^{2} - t_{\pm}(z^{\pm}) \right)
,}
where $t_{\pm}(z^{\pm})$ are integration functions determined by the
specific quantum state $|\Psi\rangle$ corresponding to the expectation value
$\langle\Psi|T^f_{\mu \nu}|\Psi\rangle \equiv \langle T^f_{\mu \nu}
\rangle$. These functions can be determined by
boundary conditions. Alternatively, Eq. {\nonlocalterms} can be obtained
by varying $NS_{PL}$. Then the functions $t_{\pm}(z^{\pm})$ arise from the
homogeneous part of the Green function in Eq. {\Polyakovaction}.
Our modified constraints (in Kruskal gauge) are
\eqn\quantumconstraints{
{\delta S_{\hbox{mod}} \over \delta g^{\pm \pm}} = 0 \; \Rightarrow \;
- \partial^{2}_{x^{\pm}}\left( e^{-2 \phi(x^{+},x^{-})} \right)
- (T^f_{\pm \pm})_{c\ell} + \kappa t_{\pm}(x^{\pm}) = 0
,}
where $(T^f_{\pm \pm})_{c\ell} =
{1\over 2} \sum_{i=1}^{N} {\left( \partial_{x^{\pm}} f_{i} \right)}^{2}$
is the classical (zero order in $\hbar$) contribution to the energy-momentum
tensor of the $f_{i}$ matter fields. $\langle T^f_{\mu \nu} \rangle$ in
{\nonlocalterms} is the one-loop quantum correction of order $\hbar$,
so the full energy-momentum tensor of the $f$-fields is
$(T^f_{\mu \nu})_{c\ell} + \langle T^f_{\mu \nu} \rangle + O(\hbar^{2})$.

For a given classical matter distribution and a given $t_{\pm}(x^{\pm})$
one finds the solution for the equations of motion
{\phirhoequations} with the constraints {\quantumconstraints}:
\eqn\quantumsolutions{
\eqalign{
e^{-2\phi} = e^{-2\rho} &= -\lambda^{2} x^{+} x^{-} -
\int^{x^{+}} dx_{2}^{+} \int^{x_{2}^{+}} dx_{1}^{+}
\left[ (T^{f}_{++})_{c\ell} - \kappa t_{+}(x_{1}^{+})
\right] \cr
&- \int^{x^{-}} dx_{2}^{-} \int^{x_{2}^{-}} dx_{1}^{-}
\left[ (T^{f}_{--})_{c\ell}
- \kappa t_{-}(x_{1}^{-}) \right] + a_{+}x^{+} + a_{-}x^{-} + b \cr }
}
where $a_{\pm}$ and $b$ are constants.
First, let us consider the linear dilaton flat space-time
solution, $e^{-2\phi}=e^{-2\rho}=
-\lambda^{2}x^{+}x^{-}$. It corresponds to the choice
$(T^{f}_{\mu \nu})_{c\ell}=0$ and
$t_{\pm}(x^{\pm})=a_{\pm}=b=0$. To determine
the corresponding quantum state $|\Psi\rangle$
one must calculate $\langle T^f_{\pm \pm} \rangle$ in
{\nonlocalterms} using the given $t_{\pm}(x^{\pm})$.
In flat coordinates $\sigma^{\pm}$,
which are related to the Kruskal coordinates $x^{\pm}$ by the
conformal coordinate transformation
$\pm \lambda x^{\pm} = e^{\pm \lambda \sigma^{\pm}}$,
the expectation values {\nonlocalterms} are
$\langle T^{f}_{\pm \pm}(\sigma^{\pm})\rangle = {\kappa \lambda^{2} \over 4}$.
We see that unlike in the RST model, in our model
$\langle T^f_{\pm \pm}(\sigma^{\pm})\rangle \neq 0$ for the linear dilaton
solution. Because $\langle T^f_{\pm \pm}\rangle =
{\kappa \lambda^{2}\over 4}$ and $\langle T^f_{+-}\rangle =0$,
the quantum state $|\Psi\rangle$ corresponding to the linear dilaton solution
may describe a system in thermal equilibrium
at temperature $T = {\lambda\over 2\pi}$.

In our model we also have
{\it static} black hole solutions.[{\ref\witten{E. Witten, {\it Phys. Rev.}
{\bf D44}, 314 (1991); S. Elizur, A. Foege and E. Rabinovici,
{\it Nucl. Phys.} {\bf B359}, 581 (1991); G. Mandal, A. Sengupta and
S. Wadia, {\it Mod. Phys. Lett.} {\bf A6}, 1685 (1991).}}]
These correspond in Eq.~{\quantumsolutions} to the choice
$(T^{f}_{\mu \nu})_{c\ell}=t_{\pm}(x^{\pm})=a_{\pm}=0$ and $b=M/\lambda$.
For these solutions at future and past null infinity (${\Im}^{+}$ and
${\Im}^{-}$, respectively) one has
$\langle T^f_{\pm \pm}\rangle = {\kappa
\lambda^{2}\over 4}$; the solutions evidently describe a black hole in
thermal equilibrium
at temperature\footnote{$^{\ddag}$}{ Since in 2D the Hawking
temperature is mass independent, one may regard the linear dilaton solution
as the zero mass limit of the static black hole solutions. This may explain
the non-zero temperature of the linear dilaton solution in our model.}
$T_{bh}={\lambda\over 2\pi}$. This is as we would expect: A static
black hole solution in a self-consistent semiclassical theory
of Hawking radiation including back-reaction is possible only if the
black hole is in thermal equilibrium with incoming radiation.

In order to find the solution corresponding asymptotically to the
Minkowski vacuum
we can use {\nonlocalterms} to find the solution for which
$\langle T^f_{\pm \pm}(\sigma^{\pm})\rangle=0$. The functions
$t_{\pm}(x^{\pm})$
are determined by imposing appropriate
boundary conditions on $\Im^{\pm}$.
We assume that on these boundaries the metric is
flat, such that $\rho(\sigma^{\pm})$ and its derivatives vanish in the
asymptotically flat coordinates $\sigma^{\pm}$. Then
the first two terms on the right-hand-side of {\nonlocalterms} vanish on the
boundary and we get
\eqn\nonlocalt{
\langle T^f_{\pm \pm}(\sigma^{\pm})\rangle\!|_{_{\hbox{boundary}}} =
- \kappa t_{\pm}(\sigma^{\pm})
,}
We see from {\nonlocalt} that the Minkowski vacuum corresponds to
$t_{\pm}(\sigma^{\pm})=0$.
To find the corresponding $t_{\pm}(x^{\pm})$ in ``Kruskal coordinates,"
one can use the tensor transformation of $\langle T^f_{\pm \pm} \rangle$ in
Eq. {\nonlocalterms} (under
a conformal coordinate transformation) and get
\eqn\trelations{
t_{\pm}(x^{\pm}) = {\left( \partial \sigma^{\pm} \over \partial x^{\pm}
\right)}^{2} \left( t_{\pm}(\sigma^{\pm}) - {1\over 2}
D^{S}_{\sigma^{\pm}}[x^{\pm}] \right) = {1\over (2x^{\pm})^{2}}
,}
where $D^{S}_{y}[z]$ is the Schwarz operator $D^{S}_{y}[z] =
\partial^{3}_{y} z / (\partial_{y} z) - {3\over 2}
{\left( \partial^{2}_{y} z / \partial_{y} z \right)}^{2}$
and we use $t_{\pm}(\sigma^{\pm})=0$.
Using {\quantumsolutions}, {\trelations} and $(T^{f}_{\mu \nu})_{c\ell}=0$,
we find that the general asymptotically Minkowski vacuum solution is
\eqn\minkowskivacuumsol{
e^{-2\phi} = e^{-2\rho} = -\lambda^{2}x^{+}x^{-} - {\kappa\over 4}
\hbox{log}(-\lambda^{2}x^{+}x^{-}) + C
,}
where $C$ is a constant.
In asymptotically flat coordinates $\sigma^{\pm} = t \pm \sigma$, we have
\eqn\staticsolution{
\eqalign{
ds^{2} &= {\left( 1 - e^{-2\lambda \sigma} (\kappa \lambda \sigma /2
- C) \right)}^{-1} (- dt^{2} + d\sigma^{2}) \cr
\phi(\sigma) &= -\lambda \sigma + \hbox{log}\left(1 - e^{-2\lambda \sigma}
({\kappa \lambda \over 2}\sigma - C) \right) .\cr }
}
This solution is static,
depending on the spatial coordinate $\sigma$ alone.
On the boundaries $\Im^{\pm}$, the solution
approaches the linear dilaton flat space-time solution, justifying
our assumption. The reason this solution with no radiation at $\Im^{\pm}$
and the earlier ones with radiation there all asymptotically approach the
linear dilaton flat space-time solution is that the coupling, $e^{2\phi}$,
of the matter to the geometry vanishes exponentially fast at $\Im^{\pm}$.

Before we turn to the question of the ground-state solution, let us consider
the ADM masses of the various solutions we have found. Suppose that
we can choose as our ground-state one of the radiationless solutions
{\staticsolution} with $C=C_{0}$, where $C_{0}$ is a constant yet to be
determined. Then the ADM mass [{\ref\Kogan{A. Bilal
and I. Kogan, {\it Phys. Rev.} {\bf D47}, 5408 (1993); S.P. de Alwis,
{\it Phys. Rev.} {\bf D46}, 5429 (1993).}},{\ref\ADM{R. Arnowitt, S. Deser
and C.W. Misner, in {\it Gravitation: An Introduction to Current Research},
ed. L. Witten (New York, Wiley, 1962); T. Regge and C. Teitelboim,
{\it Ann. Phy.} (NY), {\bf 88}, 286 (1974).}}]
of any other static solution {\staticsolution} is $\lambda(C-C_{0})$.
On the other hand, the ADM mass of the linear dilaton solution
as well as the static black hole solutions (relative
to this ground-state) is infinite. This is already clear from
the fact that these solutions have non-vanishing radiation on $\Im^{\pm}$
and can be checked explicitly by using the ADM mass definition.{\Kogan}
These considerations make it plausible that one of the static solutions
{\staticsolution} should be the ground-state. We will see later that
there exists a natural lower limit on $C_{0}$ which gives
the preferred ground-state of lowest ADM mass.

We next turn to the dynamical scenario in which the space-time is
initially described by one of the static solutions in {\staticsolution}
(not necessarily the ground state solution $C_{0}$), and in which a black hole
is formed by collapsing matter fields.
First we consider
the simple shock wave solution, but our results can be easily extended
to general infalling matter configurations.
The shock wave of infalling matter is described by
$(T^{f}_{++})_{c\ell} = {M\over \lambda x^{+}_{0}}
\delta (x^{+}-x^{+}_{0})$ and $(T^{f}_{--})_{c\ell}=0$.{\CGHS}
Unlike in the RST model, here we have a general initial
static geometry,
and the shock wave forms a black hole only if $M$,
the energy of the shock wave, is above a certain
threshold energy. We assume that $M$ is above that threshold.
Integrating $(T^{f}_{++})_{c\ell}$
in {\quantumsolutions} and using {\trelations} and $a_{\pm}=0$, we
find the evaporating black hole solution
\eqn\collapsingsolution{
e^{-2\phi} = e^{-2\rho} = -\lambda^{2} x^{+}x^{-} -{\kappa\over 4}
\hbox{log} (-\lambda^{2} x^{+}x^{-}) - {M\over \lambda x^{+}_{0}}
(x^{+} - x^{+}_{0}) \Theta (x^{+}-x^{+}_{0}) + C
,}
where $\Theta(x)$ is the standard step function.

Before the shock wave, i.e., in the region $x^{+}<x^{+}_{0}$,
we have a static solution
{\staticsolution} which is not globally flat.
If ${\kappa\over 4}[1-\hbox{log}(\kappa/4)] + C < 0$,
then the scalar curvature {\it diverges} on a timelike curve
$\sigma = \sigma_{s}$, for which $e^{-2\phi(\sigma_{s})} = 0$.
Of course this is a region of strong coupling, and one would
expect to have higher order quantum corrections there.
On the other hand, if ${\kappa\over 4}[1-\hbox{log}(\kappa/4)] + C > 0$,
the scalar curvature is bounded.
Then the region $\{x^{+}\geq 0 , x^{-}\leq 0 \}$
is geodesically incomplete and one can analytically extend it
to $x^{-}>0$ and $x^{+}<0$. Also in this case there is a region of strong
coupling near $\sigma = \sigma_{min} = -{1\over2\lambda}
\hbox{log}({\kappa\over 4})$.
In the semiclassical approximation, one avoids the strong coupling region
by imposing boundary conditions on a suitable time-like
hypersurface.[{\ref\Verlinde{K. Schoutens,
E. Verlinde and H. Verlinde, {\it Phys. Rev.} {\bf D48}, 2670
(1993).}},14,{\ref\Klebanov{C. Callan, I. Klebanov,
A. Ludwig and J. Mandacena, ``Exact Solutions of a Boundary Conformal
Field Theory", Princeton preprint no. PUPT-1450, hepth@xxx/9402113 (1994);
J. Polchinski and L. Thorlacius, {\it Phys. Rev.} {\bf D50}, 5177 (1994);
A. Strominger and L. Thorlacius, ``Conformally Invariant Boundary
Conditions for Dilaton Gravity", UCSB preprint no NSF-ITP-94-34 (1994).}}]
For the static solutions {\staticsolution},
$\langle T^f_{++}(\sigma^{\pm})\rangle$ and $\langle
T^f_{--}(\sigma^{\pm})\rangle$ are constant
on {\it any} time-like hypersurface $\sigma=\hbox{const}$. Moreover
\eqn\equalt{
\langle T^f_{++}(\sigma^{\pm})\rangle\!|_{_{\sigma=\sigma_{0}}} = \;
\langle T^f_{--}(\sigma^{\pm})\rangle\!|_{_{\sigma=\sigma_{0}}} \qquad
\hbox{for any constant $\sigma_{0}$}
.}
This means that we can limit our model to a region in which the
semiclassical approximation is valid by imposing reflecting boundary
conditions {\equalt} on any time-like hypersurface $\sigma=\sigma_{0}$ that
lies
outside the region of strong coupling (these boundary conditions are also
conformal {\Klebanov}).
The geometry before the shock wave is therefore a static geometry,
defined in the region $\sigma > \sigma_{s}$ in the case
${\kappa\over 4}[1-\hbox{log}(\kappa/4)] + C < 0$
(or $\sigma > \sigma_{min}$ in the case
${\kappa\over 4}[1-\hbox{log}(\kappa/4)] + C > 0$),
with reflecting boundary conditions on $\sigma=\sigma_{s}+\delta$
(or on $\sigma=\sigma_{min}+\delta$) where $\delta$ is an
arbitrary small positive constant.

The solution to the future of the shock
wave ($x^{+}>x^{+}_{0}$) is (see {\collapsingsolution})
\eqn\bhsolution{
e^{-2\phi} = e^{-2\rho} = -\lambda^{2} x^{+} (x^{-} + \Delta)
- {\kappa\over 4} \hbox{log}(-\lambda^{2}x^{+}x^{-}) + {M\over \lambda}
+ C
,}
where $\Delta = {M\over \lambda^{3}x^{+}_{0}}$. This solution is
asymptotically flat and describes a black hole with a
singularity at $e^{-2\phi}=0$. The black hole singularity curve is
\eqn\singularity{
 -\lambda^{2} x_{s}^{+} (x_{s}^{-} + \Delta)
- {\kappa\over 4} \hbox{log}(-\lambda^{2}x_{s}^{+}x_{s}^{-})
+ {M\over \lambda} + C = 0
.}
Initially the singularity is behind an apparent horizon
$\partial_{+}e^{-2\phi} = 0$,[{\ref\rusuth{J.R. Russo, L. Susskind and
L. Thorlacius, {\it Phys. Lett.} {\bf B292}, 13 (1992).}},{\yoav}]
which is the curve
\eqn\apparenthorizon{
-\lambda^{2}x^{+}_{h}(x^{-}_{h} + \Delta) = {\kappa\over 4}
.}
When the apparent horizon is formed, the black hole starts radiating.
One can see this by calculating $\langle T^f_{\mu \nu}\rangle$ at future null
infinity
($x^{+} \rightarrow \infty$). From {\bhsolution} we see that the
asymptotically flat coordinates on ${\Im}^{+}$ are
${\widehat{\sigma}}^{\pm}$, related to $x^{\pm}$ by the conformal
coordinate transformation,
$\lambda {\widehat{\sigma}}^{+} = \hbox{log}(\lambda x^{+})$
and $-\lambda {\widehat{\sigma}}^{-} = \hbox{log}( -\lambda (x^{-}+\Delta))$.
Using {\nonlocalt} and {\trelations} we get
\eqn\hawkingradiation{
\langle T^f_{--}({\widehat{\sigma}}^{\pm})\rangle\!|_{_{{\Im}^{+}}} =
{\kappa\lambda^{2}\over 4} \left( 1 - {1\over (1 +
\lambda \Delta e^{\lambda {\hat{\sigma}}^{-}})^2} \right)
.}
This is the ``standard" Hawking radiation in 2D, where the
Hawking temperature $T_{H}={\lambda\over 2\pi}$ is a constant.{\CGHS}
One can further verify that when the black hole
evaporates over a long period of time, i.e., if $M >> \kappa \lambda$,
the spectrum of the Hawking radiation is indeed
Planckian.[{\ParkerWald},{\ref\Nelson{S.B. Giddings and W.M. Nelson,
{\it Phys. Rev.} {\bf D46}, 2486 (1992).}}]

As the black hole evaporates by emitting Hawking radiation, the
apparent horizon shrinks and eventually
meets the singularity in a {\it finite}
proper time. They intersect at (see Fig. 1)
\eqn\nakedsingularity{
x^{+}_{int} = {1\over \lambda^{2}\Delta} \left(
e ^{({4(M+\lambda C)\over \kappa \lambda}
+ 1)} - {\kappa\over 4} \right) \qquad \hbox{and} \qquad
x^{-}_{int} = - \Delta {\left( 1 - {\kappa\over 4}
e^{-({4(M+\lambda C)\over \kappa \lambda} + 1)} \right)}^{-1}
.}
At this point the singularity becomes naked. We show that it is possible
to impose a boundary condition in which a weak shock wave emanates from
the intersection point, resulting in a solution that is stable (having
non-negative ADM mass), conserves energy, and is continuous with the
metric defined to the past of the null hypersurface $x^{-}=x^{-}_{int}$.


Before considering the solution to the future of the null hypersurface
$x^{-}=x^{-}_{int}$ (the end-state solution), we calculate the total
amount $E_{rad}$ of energy radiated during the evaporation.
Integrating {\hawkingradiation} over ${\Im}^{+}$
(up to $x^{-}_{int}$) gives
\eqn\radiationenergy{
E_{rad} =
\int_{-\infty}^{{\hat{\sigma}}^{-}_{int}}
\langle T^f_{--}({\widehat{\sigma}}^{-})\rangle d{\widehat{\sigma}}^{-} = M
+\lambda C -
{\kappa \lambda \over 4} \left( \hbox{log}(\kappa/4) - 1 \right)
-{\kappa \lambda \Delta \over 4 x^{-}_{int}}
,}
where ${\widehat{\sigma}}^{-}_{int}={\widehat{\sigma}}^{-}(x^{-}_{int})$.
The result {\radiationenergy} is {\it exact}.
The ADM mass {\Kogan} of the dynamical solution
{\collapsingsolution} (relative to the ground state $C=C_{0}$)
is $M_{ADM} = M + \lambda (C-C_{0})$.
We see that the black hole radiates
almost all of its initial energy.
The unradiated mass $\delta M$ remaining as $x^{-} \rightarrow x^{-}_{int}$
(which is the Bondi mass) is
\eqn\remainingmass{
\delta M = M_{ADM} - E_{rad} = {\kappa \lambda\over 4}
\left( \hbox{log}(\kappa/4) - 1 \right) - \lambda C_{0}
+ {\kappa \lambda \Delta \over 4 x^{-}_{int}}
.}

We now consider the solution to the future of the point of intersection
$(x^{+}_{int},x^{-}_{int})$.
A natural candidate for such an end-state
in our model is one of the static solutions {\staticsolution},
so we try to find boundary conditions such that
the solution {\bhsolution} is continuously matched to
one of the static solutions {\staticsolution}. Remember that the
asymptotically flat coordinates are ${\widehat{\sigma}}^{\pm}$,
so one should replace $\sigma$
in {\staticsolution} with $\widehat{\sigma}={1\over 2}(\widehat{\sigma}^+
- \widehat{\sigma}^- )$.
In the $x^{\pm}$ coordinates the corresponding
static solution is (see {\minkowskivacuumsol})
\eqn\staticfuturesol{
e^{-2\phi}=e^{-2\rho}=-\lambda^{2}x^{+}(x^{-}+\Delta) - {\kappa\over 4}
\hbox{log}(-\lambda^{2} x^{+}(x^{-}+\Delta)) + \widehat{C}
.}
We would like to see if there exists a constant $\widehat{C}=C^{*}$,
such that on the null hypersurface
$x^{-}=x^{-}_{int}$ the solutions {\bhsolution} and {\staticfuturesol}
can be matched continuously. This is indeed the case and from
{\nakedsingularity}, {\bhsolution} and {\staticfuturesol} we get
$C^{*} = -{\kappa\over 4}(1-\hbox{log}(\kappa/4))$. The end-state
solution, or ``remnant", is therefore
\eqn\endstatesolution{
e^{-2\phi}=e^{-2\rho}=-\lambda^{2}x^{+}(x^{-}+\Delta) - {\kappa\over 4}
\hbox{log}(-\lambda^{2} x^{+}(x^{-}+\Delta)) - {\kappa\over 4} (1 -
\hbox{log}(\kappa/4))
,}
where $x^{-} > x^{-}_{int}$.
{}From the constraint equations {\quantumconstraints} we find that
\eqn\thunderpop{
(T^f_{--}(\widehat{\sigma}^{-}))_{c\ell} = {1\over 2} \sum_{i=1}^{N}
(\partial_{-} f_{i})^{2} = {\kappa \lambda \Delta\over
4x^{-}_{int}}  \delta ({\widehat{\sigma}}^{-}-{\widehat{\sigma}}^{-}_{int})
.}
This describes a shock wave originating at the intersection point and
carrying a small amount of negative
energy, $\kappa \lambda \Delta / (4x^{-}_{int})$,
to null infinity. One may call it a ``thunderpop".{\RST} The solution
{\endstatesolution} is one of the static solutions
that is asymptotically flat (with no radiation) on ${\Im}^{+}$.
This means that there is no Hawking radiation after
the thunderpop {\thunderpop}.

The mass remaining after the shockwave {\thunderpop} has been emitted
is  $\delta M - \kappa \lambda \Delta / (4 x^{-}_{int})$. One readily
verifies that this is
equal to the mass of the ``remnant" (relative to $C_{0}$)
$M_{rem}=\lambda(C^{*}-C_{0})$. The fact that energy is
exactly conserved, including terms of order $\hbar$, supports the
self-consistency of our semi-classical theory.
Notice that $C^{*}$ and therefore the ``remnant" mass is
independent of the mass $M$ of the infalling matter and of the constant $C$
describing the initial static geometry.
Even more surprising is the fact that the end-state solution with
$\widehat{C}=C^{*}$
is the critical solution separating singular and non-singular static
solutions described by Eq.~{\staticfuturesol}.
For $\widehat{C}>C^{*}$ the curvature of the solution {\staticfuturesol} is
bounded, while for $\widehat{C}<C^{*}$ the curvature diverges on a time-like
curve $\widehat{\sigma}=\widehat{\sigma}_{s}$, for which
$e^{-2\phi(\hat{\sigma}_{s})}=0$.
In solution space, the solution {\endstatesolution} that has
$\widehat{C}=C^{*}$ is the
boundary between these two different classes of solutions.

Consider the late-time space-like hypersurface $\Sigma$ shown in Fig. 1.
Its right boundary ($\widehat{\sigma} \rightarrow \infty$) is $i^{0}$,
while its left boundary is the curve $\widehat{\sigma} =
\widehat{\sigma}_{cr}$, for which
$e^{-2\phi}=0$. For the critical solution we have
$\partial_{x^{+}}(e^{-2\phi(\hat{\sigma}_{cr})}) =
e^{-2\phi(\hat{\sigma}_{cr})}=0$ and
the curve $\widehat{\sigma} = \widehat{\sigma}_{cr}$ is the analytical
continuation of the apparent horizon to the region $x^{-} > x^{-}_{int}$.
We define $\epsilon \equiv \widehat{\sigma}
- \widehat{\sigma}_{cr}$, and calculate the metric near $\epsilon = 0$.
{}From {\endstatesolution} we get
\eqn\throatsolution{
ds^{2} \rightarrow {-d{\hat{t}}^{2} + d{\epsilon}^{2}\over
2\lambda^{2} \epsilon^{2} + {\cal O}(\epsilon^{3})}
,}
where $\widehat{t} = {1\over 2} (\widehat{\sigma}^{+}+\widehat{\sigma}^-)$.
An important feature of {\throatsolution} is that there is no
linear term (in $\epsilon$) in its denominator. The first non-vanishing term
is of order $\epsilon^{2}$, which means that the geometric structure near
$\epsilon=0$ is that of an {\it infinite throat}.
Consider for example the distance
along $\widehat{t}=\hbox{constant}$ curves. The distance to $\widehat{\sigma}=
\widehat{\sigma}_{cr}$ diverges logarithmically, exactly as it does
in higher-dimensional extremal black holes. The end-state space-time
is geodesically complete. One may consider this solution as an
``extremal 2D black hole". On $\widehat{\sigma}=\widehat{\sigma}_{cr}$
the Ricci scalar is constant,
$R^{(2)}=4\lambda^{2}$
and the geometry is regular.

The most natural choice of $C_{0}$ for ground-state
solution is the one with $C_{0}=C^{*}$.
This solution describes
a static radiationless geometry which is regular everywhere.
Any solution {\staticsolution} with smaller ADM mass
($C<C^{*}$) has a naked singularity. In the class of solutions with no
naked singularities, $C=C^{*}$ is the one with
lowest energy. This is very similar to
the linear dilaton vacuum solution (LDV) in classical dilaton gravity
or to Minkowski space in Einstein gravity.\footnote{$^{\S}$}{Classical
solutions with ADM mass smaller than the LDV
have a naked singularity, as do Schwarzschild solutions with mass
smaller than zero.}
Also if we choose $C_{0}=C^{*}$, then the mass remaining
after the thunderpop {\thunderpop} is {\it exactly} zero. Thus the
end-state solution {\endstatesolution} is the static ground-state.
Its geometrical structure is {\it independent} of the initial conditions and
is a semi-infinite throat extending into the strong coupling region.

In our 2D semiclassical model, one does not recover all
the information of the initial state from the end-state solution.
For infalling matter described by a general $(T^{f}_{++})_{c\ell}$
of compact support,
the solution {\quantumsolutions} will
depend only on the first two
moments of $(T^f_{++})_{c\ell}$, $M=\lambda\int x^{+} (T^f_{++})_{_{c\ell}}
dx^{+}$ and $P_{+}=\int (T^f_{++})_{_{c\ell}}dx^{+}$.{\RST}
The end-state solution will still be
{\endstatesolution}, but with $\Delta=\lambda^{-2}P_{+}$. The information
encoded in this ``remnant" (or more precisely, in its past null
boundary $x^{-}=x^{-}_{int}$) is only about $P_{+}$ and $M$.
Thus in our semiclassical model this end-state solution does not
qualify as the ``cornucopion" of Ref. [{\ref\banks{T.
Banks, A. Dabolkhar,
M. Douglas and M. O'Loughlin, {\it Phys. rev.} {\bf D45}, 3607 (1992);
T. Banks M. O'Loughlin and A. Strominger, {\it Phys. Rev.} {\bf D47},
4476 (1993); T. Banks, ``Lectures on Black holes and Information Loss",
Rutgers Univ. preprint no. RU-94-91, hepth@xxx/9412131 (1994).}}].
However, the semi-infinite throat extends to a region of very
strong coupling. There may be sufficient freedom in
this strong coupling region to encode more information through strong
quantum gravitational effects.

In this work we constructed an action in 2D dilaton gravity and showed
that, with a natural boundary condition, all evaporating black holes
in our model
end in a unique ground-state geometry having a semi-infinite throat.

\bigbreak\bigskip\bigskip\centerline{\bf Acknowledgements}\nobreak
We thank Bruce Allen, John Friedman, and Jorma Louko,
for helpful discussions and the National Science Foundation for
support under grant PHY-9105935.

\baselineskip=30pt
\listrefs
\end